\documentclass[aps,pra,reprint,a4paper,showpacs]{revtex4-1}
\usepackage{amsmath}
\usepackage[latin9]{inputenc}
\usepackage{graphicx}
\usepackage{hyperref}
\usepackage{siunitx}
\usepackage{epstopdf}
\usepackage{graphicx}
\usepackage{todonotes}
\usepackage[normalem]{ulem}
\usepackage{color}

\DeclareSIUnit\gauss{G}

\begin{document}

\title{Controlling the magnetic field sensitivity of atomic clock states by microwave dressing}

\author{L. S\'ark\'any}
\author{P. Weiss}
\author{H. Hattermann}\email{hattermann@pit.physik.uni-tuebingen.de}
\author{J. Fort\'agh}
\affiliation{CQ Center for Collective Quantum Phenomena and their
Applications, Physikalisches Institut, Eberhard Karls Universit\"at
T\"ubingen, Auf der Morgenstelle 14, D-72076 T\"ubingen,
Germany}
\begin{abstract}
We demonstrate control of the differential Zeeman shift between clock states of ultracold rubidium atoms by means of non-resonant microwave dressing. 
Using the dc-field dependence of the microwave detuning, we suppress the first and second order differential Zeeman shift in magnetically trapped $^{87}$Rb atoms.
By dressing the state pair 5S$_{1/2} F= 1, m_F = -1$ and $F= 2, m_F = 1$, a residual frequency spread of ${<}0.1$\,Hz in a range of 100\,mG around a chosen magnetic offset field can be achieved. 
This is one order of magnitude smaller than the shift of the bare states at the magic field of the Breit-Rabi parabola.
We further identify \textit{double magic} points, around which the clock frequency is insensitive to fluctuations both in the magnetic field and the dressing Rabi frequency.
The technique is compatible with chip-based cold atom systems and allows the creation of clock and qubit states with reduced sensitivity to magnetic field noise.
\end{abstract}
\maketitle
\section{Introduction}
The sensitivity of atomic transitions to external field perturbations represents a major limitation for the accuracy and stability of atomic clocks \cite{Bloom2014, Ushijima2014} and for the time of quantum information storage in ultracold atoms and atomic gases \cite{Zhao2009c, Bao2012, Dudin2013}. 
Electromagnetic field fluctuations and inhomogeneous trapping potentials give rise to temporal and spatial variations of atomic transition frequency. 
A common approach to reduce the frequency broadening observed in the preparation and read out of atomic superposition states is the use of `magic' magnetic fields \cite{Harber2002, Derevianko2010a} and wavelengths \cite{Katori2003, Katori2011, Chicireanu2011} for which the differential Zeeman  and Stark shift of a state pair is minimized, respectively.
In addition, density dependent collisional shifts \cite{Treutlein2004, Rosenbusch2009} and the effect of identical spin rotation \cite{Deutsch2010, Kleine2011, Bernon2013} have been shown to counteract inhomogeneous dephasing of superposition states in trapped atomic clouds. 
Possible realizations of cold atom quantum memories on atom chips \cite{Verdu2009, Patton2013, Patton2013a, Bernon2013, Xiang2013} and chip based atomic clocks \cite{Rosenbusch2009, Farkas2010, Vuletic2011, Ramirez2011} have to face additional perturbations due to the proximate solid surface \cite{Fortagh07}. 
Controlling differential shifts between clock states is one of the key requirements for the realization of a coherent interface between cold atoms and solid-state quantum electronic circuits \cite{Rabl2006, Wallquist2009, Hafezi2012, Petrosyan2008, Petrosyan2009, Braun2011, Gao2011, Henschel2010, Dora2013}. 
It was recently shown that Rydberg states can be rendered insensitive to small variations of electric fields by microwave dressing \cite{Jones2013}. 
Previously, radio-frequency dressing of nuclear spins has been proposed to cancel differential Zeeman shifts between optical clock transitions \cite{Zanon2012}.
Microwave fields have further been used to suppress the magnetic field dependency of qubit states in trapped ions \cite{Timoney2011} and nitrogen-vacancy centers with both pulsed \cite{Lange2010, Souza2011} and continuous \cite{Xu2012, Cai2012} decoupling schemes.

Here, we demonstrate the control and suppression of the differential Zeeman shift between atomic qubit states up to second order by microwave dressing, thereby reducing the magnetic field sensitivity of the clock transition frequency.
The technique does not require the state pair to be close to magic magnetic fields but can be applied for a wide range of chosen magnetic offset fields in the trap.
By dressing the state pair  5S$_{1/2} F= 1, m_F = -1$ and $F= 2, m_F = 1$ of $^{87}$Rb, we demonstrate that a variation of ${<}0.1$\,Hz over a magnetic field range of ${>}100$\,mG can be achieved, one order of magnitude less than the differential shift of the bare states around the magic offset field. 
In addition, we demonstrate the existence of points where this dressing becomes insensitive to fluctuations in the dressing Rabi frequency, enabling the generation of noise protected qubit states.
Our model and experimental results show that the frequency of an atomic clock can be engineered by microwave dressing to achieve arbitrary curvatures, e.g. nearly zero differential shift, around a given magnetic offset field.

\section{Microwave dressing of atomic transitions}

\begin{figure*}
\centerline{\includegraphics[width=.95\textwidth]{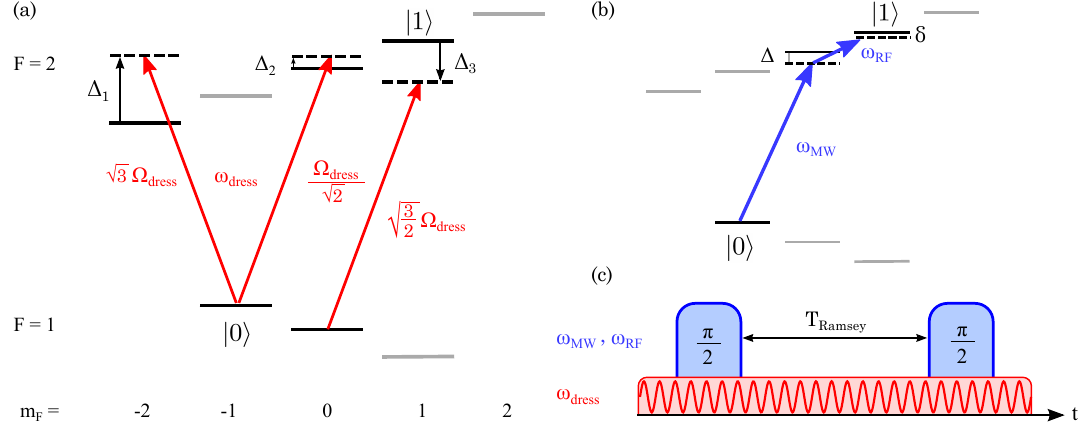}}
\caption{(a) Ground state hyperfine structure and Zeeman sublevels of $^{87}$Rb in a magnetic field. A microwave field of frequency $\omega_\text{dress}$ and Rabi frequency $\Omega_\text{dress}$ is used for dressing the clock transition. (b) The atomic transition is probed by means of Ramsey interferometry. A two-photon pulse with $\omega_\text{MW} \approx 6.833$\,GHz and $\omega_\text{RF} \approx 2$\,MHz is used to drive the transition. (c) Schematic of the experimental sequence. A $\pi/2$-pulse is used to prepare the atoms in a superposition state $1/\sqrt{2}\cdot( \left|0\right\rangle + \left|1\right\rangle)$. After a variable hold time $T_\text{Ramsey}$, the interferometer is closed by the application of a second $\pi/2$ pulse and the population of the two states, oscillating with frequency $\delta$, is measured. The dressing field is  left on throughout the interferometer sequence.}
\label{fig:level_scheme}
\end{figure*}

In static magnetic fields, the degeneracy of the hyperfine levels of ground state alkali atoms is lifted according to the Breit-Rabi-formula \cite{Breit1931}.
In small and intermediate fields, the Zeeman effect can be expanded in terms linear and quadratic in the magnetic field strength.
The interaction of the atom with non-resonant ac electromagnetic fields leads to the ac-shift  of the levels, which depends on the detuning, e.g. $\Delta E \propto \Omega_\text{dress}^2/\Delta_\text{dress}$.
As this detuning depends on the dc-Zeeman shift of the levels, a suitable choice of the microwave field allows compensating spatial and temporal variations of the differential Zeeman shift.

While the technique presented here can be applied for all alkali elements, we now discuss this for the specific case of $^{87}$Rb.
The two photon transition 5S$_{1/2} F= 1, m_F = -1 \rightarrow F= 2, m_F = 1$ is commonly used as atomic clock transition for magnetically trapped $^{87}$Rb.
Both states exhibit nearly the same first-order Zeeman shift, starkly reducing the sensitivity of the transition to magnetic field fluctuations and making the two states ideal candidates as atomic qubit states. 
The energy of the two states $\left|0\right\rangle\ \equiv $ 5S$_{1/2} F= 1, m_F = -1$ and $\left|1\right\rangle \equiv $ 5S$_{1/2} F= 2, m_F = 1$ in a magnetic field of magnitude $B$ is given by
\begin{equation}
    E_0 = \mu_1B - 3\beta B^2, \textrm{\ and}
    \label{eq:E_lower_0}
\end{equation}
\begin{equation}
    E_1 = \mu_2B + 3\beta B^2 +\omega_0,
    \label{eq:E_upper_0}
\end{equation}
where $\mu_1 =2\pi\cdot 702.37$\,kHz/G, $\mu_2 = 2\pi\cdot699.58$\,kHz/G, $\beta =2\pi\cdot 71.89$\,Hz/G$^2$, and $\omega_0 = 2\pi\cdot 6.8346826109$\,GHz \cite{Steck2010} is the frequency difference of the two states in the absence of any fields.
The energy difference between the two levels can be expressed by 
\begin{equation}
\Delta E_{0,1}/\hbar = 6 \beta\cdot(B-B_0)^2 + 2\pi\cdot 6.8346781136\,\mathrm{GHz},
\end{equation}
where $B_0\approx 3.229$\,G is the so-called magic offset field \cite{Lewandowski2002}.
Using microwave dressing of the Zeeman sublevels with an appropriate frequency  $\omega_\text{dress}$ and Rabi-frequency $\Omega_\text{dress}$, the second-order Zeeman shift can be compensated for.

The microwave field leads to a correction of the form 
\begin{equation}
\Delta E_{\text{dress},i} = \sum_{i,\alpha}\frac{\Omega_{i,\alpha}^2}{\Delta_{i,\alpha}}
\end{equation}
for both of the states $i$, where $\alpha  = \sigma_+, \sigma_-, \pi$ denotes all the possible polarizations of the dressing field, for each of which the relevant detuning $\Delta_\alpha$ and Rabi frequency $\Omega_\alpha$ needs to be taken into account.
If we consider a microwave field $B_\text{dress}\cdot \cos(\omega_\text{dress}t)$, which is linearly polarized perpendicular to the quantization axis (given by the magnetic offset field), the situation is simplified, as we only need to take into account $\sigma_+$ and $\sigma_-$ transitions, as sketched in Fig.\ \ref{fig:level_scheme}(a).

\begin{figure}
\centerline{\includegraphics[width=.5\textwidth]{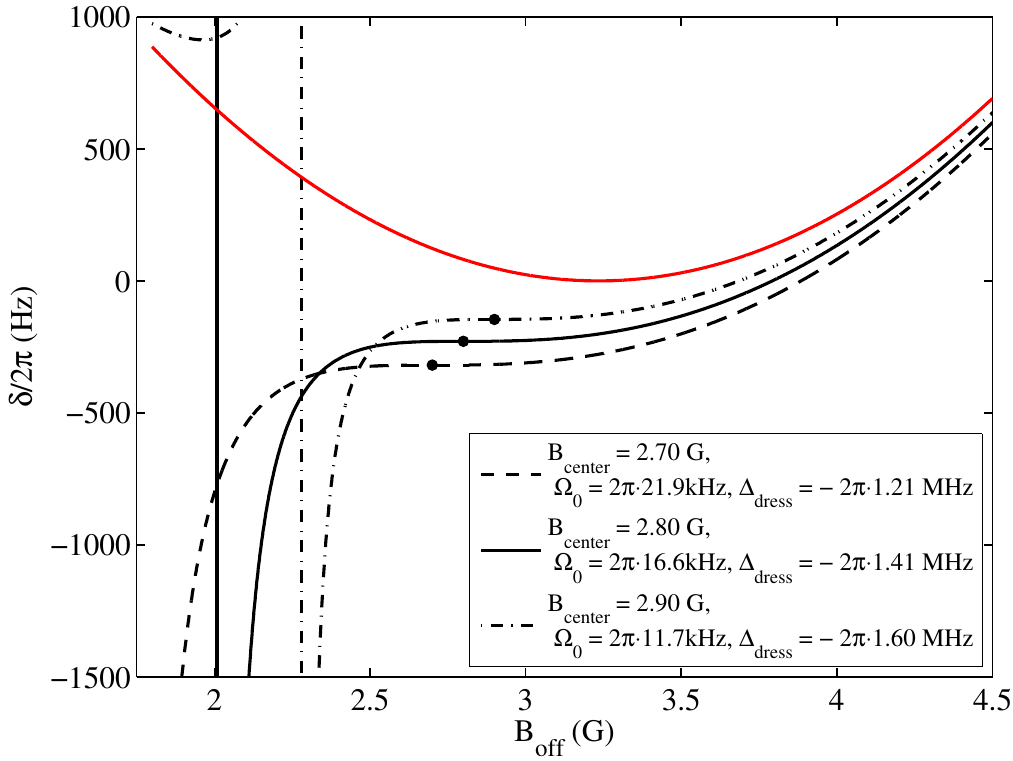}}
\caption{Calculated frequency difference of the clock transition as a function of the magnetic field. The Breit-Rabi parabola for the case without the dressing field is plotted in red. The three black curves show the cancellation of the $B_\text{off}$-dependence around three different central values $B_\text{center}$. For an arbitrary value of $B_\text{center}$, the optimal detuning and Rabi-frequency can be calculated.}
\label{fig:freq_v_b}
\end{figure}

In the rotating wave approximation, the Hamiltionian relevant for the two states is

\begin{equation}
 H_0 = \hbar \cdot  \left[
        \begin{array}{ccc}
            0 & \sqrt 3 \Omega_{\text{dress}} & \frac{1}{\sqrt2}\Omega_{\text{dress}} \\
            \sqrt3 \Omega_{\text{dress}} & -\Delta_1 & 0 \\
            \frac{1}{\sqrt2}\Omega_{\text{dress}} & 0 & -\Delta_2
        \end{array}
    \right] \textrm{, and}
    \label{eq:Heff_MWpm}
\end{equation}

\begin{equation}
H_1 = \hbar \cdot  \left[
        \begin{array}{cc}
            0 & \sqrt{\frac{3}{2}} \Omega_{\text{dress}} \\
            \sqrt{\frac{3}{2}} \Omega_{\text{dress}} & -\Delta_3
        \end{array}
    \right],
    \label{eq:Heff_MWpm_upper}
\end{equation}
where we have defined $\Omega_{\text{dress}}$ as state-independent Rabi frequency
\begin{equation}
    \Omega_\text{dress} = \frac{1}{2\sqrt{2} \hbar}\, \mu_B\, g_F\, \left| B_\text{dress}\right|, 
    \label{eq:Rabi_convetion}
\end{equation}
and the values of the detunings in Eqs.\ \ref{eq:Heff_MWpm},\ref{eq:Heff_MWpm_upper} and Fig.\ \ref{fig:level_scheme}(a) are given by
\begin{equation}
    \Delta_{1} =  {\Delta_\text{dress} +(\mu_1+2\mu_2)B-3\beta B^2},
    \label{eq:delta_sigmaM}
\end{equation}
\begin{equation}
    \Delta_{2} = {\Delta_\text{dress} +\mu_1B-7\beta B^2}, \mathrm{\ and}
    \label{eq:delta_sigmaP}
\end{equation}
\begin{equation}
    \Delta_{3} = {\Delta_\text{dress} - \mu_2B - 7\beta B^2}, 
    \label{eq:delta_pi}
\end{equation}
where $\Delta_\text{dress} = \omega_\text{dress} - \omega_0$.
With this notation, we can write the frequency difference between the two states as
\begin{equation}
\begin{split}
\Delta E_{0,1}/\hbar = \omega_0 + (\mu_2-\mu_1)B + 6\beta B^2 -\dots\\- \Omega_\text{dress}^{2} \left(  \frac{3}{\Delta_1(B)} + \frac{1/2}{\Delta_2(B)} +\frac{3/2}{\Delta_3(B)} \right).
\end{split}
\label{eq:diff}
\end{equation}
For a given offset field, it is now possible to find numerical solutions $\omega_\text{dress}$ and $\Omega_\text{dress}$ for which the first and second derivatives of Eq.\ (\ref{eq:diff}) with respect to $B$ disappear, i.e. the Zeeman shift of the transitions around that offset field is canceled up to second order. 
This is illustrated in Fig.\ \ref{fig:freq_v_b}, where the frequency difference is plotted as a function of the magnetic field for different values of the center of the plateau $B_\text{center}$. 
For each curve, different optimized values for $\omega_\text{dress}$ and $\Omega_\text{dress}$ were calculated.
The choice of $B_\text{center}$ is completely arbitrary within the limits of Rabi frequencies $\Omega_\text{dress}$ that are achievable in experimental conditions. 

\section{Experimental procedure}
The measurements are taken with atomic clouds magnetically trapped on a superconducting atom chip. 
Atoms are loaded into this trap as follows  \cite{Bernon2013}:
an ensemble of ultracold $^{87}$Rb atoms is prepared in a magneto-optical trap and subsequently transferred into a Ioffe-Pritchard-type magnetic trap situated in the room-temperature environment of our setup \cite{Cano2011}.
The atomic cloud is cooled by forced radio-frequency evaporation and then loaded into an optical dipole trap used to transport the ensemble to a position below the superconducting atom chip at 4.2\,K.
We load an ensemble of ${\sim} 10^6$ atoms at a temperature of ${\sim} \SI{1}{\micro \kelvin}$ into the magnetic chip trap, which is based on a $Z$-wire geometry \cite{Fortagh07}. 
The oscillation frequencies in the trap are given by $\omega_x = 2\pi\cdot 30$\,s$^{-1}$, $\omega_y = 2\pi\cdot 158$\,s$^{-1}$, $\omega_z = 2\pi\cdot 155$\,s$^{-1}$ and the offset field $B_\text{off}$, which defines the the quantization axis, is pointing along the $x$-direction.
The atomic cloud in the magnetic trap is cooled to a temperature of ${\sim} 250$\,nK by evaporation.
After this sequence, which is repeated every ${\sim} 23$\,s, we end up with an ensemble of roughly $10^5$ atoms.
After a holdtime of 2\,s in the magnetic trap, which allows for damping of possible eddy currents in the metallic chip holder, a microwave field for dressing is applied.

The microwave field is irradiated from an antenna outside of the vacuum chamber and is counterpropagating to the quantization axis.
We measured the polarization of the microwave by driving resonant $\sigma_+$ and $\sigma_-$ Rabi oscillations.
We found a ratio of $\sqrt{6}\Omega_{0,\sigma +}/\Omega_{0,\sigma -} \approx 0.81$, while for a linear (circular) polarization the expected ratio would be 1 (0).
The factor $\sqrt{6}$ stems from the different transition strengths, as visible in the Hamiltonian in Eq.\ (\ref{eq:Heff_MWpm}).

The frequency of the transition is measured by means of Ramsey interferometry. 
The interferometric sequence is started 100\,ms after switching on the dressing field by applying a combined microwave and radio-frequency two-photon pulse with a pulse area of $\pi/2$ ( $ T_{\pi/2} = \SI{137}{\micro\second} $), which prepares the atomic ensemble in a coherent superposition of states $\left|0\right\rangle$ and $\left|1\right\rangle$, see Fig.\ \ref{fig:level_scheme}(b) and (c).
The microwave pulses are irradiated from a second external antenna with a wave vector  perpendicular to the quantization axis, while the radio-frequency field is generated by an alternating current in the trapping wire.
Both frequencies are chosen with a detuning of $\Delta \sim 2\pi\cdot  310$\,kHz with respect to the transition to the intermediate level 5S$_{1/2} F= 2, m_F = 0$, so that the probability of populating this level is negligible.
After a variable holdtime $T_\text{Ramsey}$, the interferometer is closed by a second $\pi/2$-pulse and we measure the population of the two states $\left|0\right\rangle$ and $\left|1\right\rangle$, which oscillates with the angular frequency $\delta = |\omega_\text{MW} + \omega_\text{RF} - \Delta E_{0,1}/\hbar|$.
We determine this frequency $\delta$ for different offset fields $B_\text{off}$ and seek to eliminate the magnetic field dependence of the transition.

\section{Measurements \& Discussion}
\begin{figure}
\centerline{\includegraphics[width=.5\textwidth]{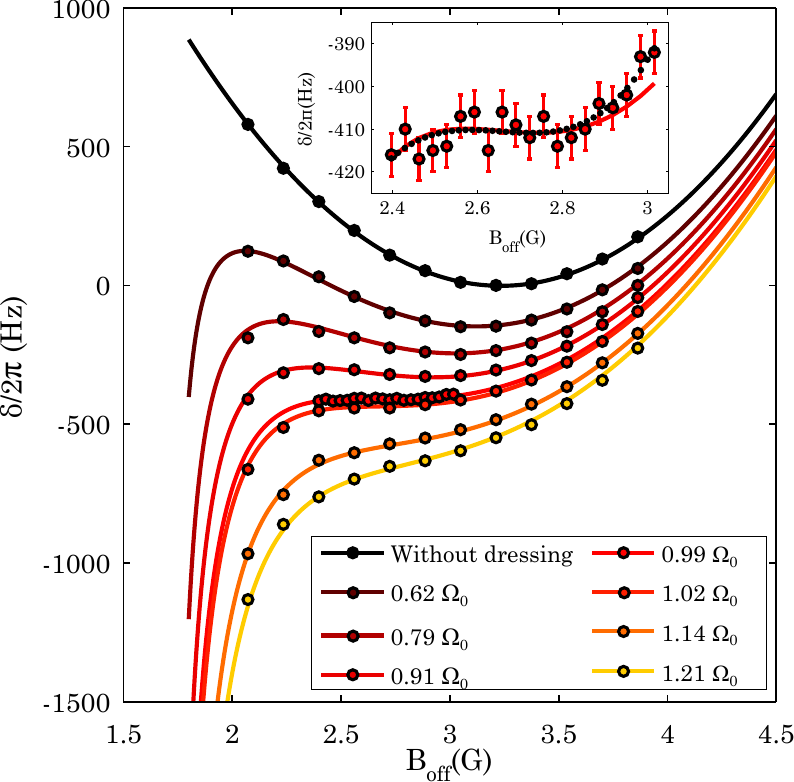}}
\caption{Measurement of the differential Zeeman shift between the states $\left|0\right\rangle$ and $\left|1\right\rangle$ for different Rabi frequencies. The frequency zero-point was set to the frequency at the magic offset field without dressing. For a Rabi frequency $\Omega_\text{dress}= \Omega_0 = 2\pi\cdot 20.1$\,kHz, the frequency is nearly independent of the magnetic offset field in a range of $\pm 100$\,mG around the chosen value $B_\text{center} = 2.65$\,G. \textit{Inset}: Detail of the curve with $\Omega_\text{dress} = 0.99\cdot \Omega_0$. We estimate a measurement error of $\pm 5$\,Hz resulting from fluctuations of the MW power. The theory curve (solid red) is plot along a polynomial fit (dotted black), showing the suppression of the first and second order Zeeman shift down to a level of -7.3\,Hz/G and 5.0\,Hz/G$^2$.  }
\label{fig:results}
\end{figure}
To demonstrate the control over the differential Zeeman shift, we measure the frequency of the Ramsey interferometer as a function of the magnetic offset field $B_\text{off}$ for different powers of the dressing field (Fig.\ \ref{fig:results}). 
For each value of $B_\text{off}$, we adjust $\omega_\text{RF}$ and $\omega_\text{MW}$ in order to keep the detuning $\Delta$ to the intermediate state constant, while keeping the sum frequency $\omega_\text{RF} + \omega_\text{MW}$ fixed.
The measurement without dressing field yields the expected Breit-Rabi parabola which we use to calibrate the magnetic field $B_\text{off}$.

For the cancellation of the magnetic field dependence, a magnetic offset field  $B_\text{off} = 2.65$\,G was chosen.
For this $B_\text{off}$, we calculated the optimum detuning $\Delta_\text{dress}$ and Rabi-frequency $\Omega_\text{dress} $ for the measured ratio between $\sigma_+$ and $\sigma_-$ transition strengths. 
We  measure $\delta$ vs.\ $B_\text{off}$ in the range of 2.1\,G to 3.8\,G for Rabi-frequencies in the range of $2\pi\cdot 12$\,kHz to $2\pi\cdot 25$\,kHz with a calculated optimal Rabi frequency $\Omega_0 = 2\pi\cdot20.1$\,kHz.
The results of these measurements are plotted in Fig.\ \ref{fig:results} along with the results of the analytical calculations, taking into account the measured imbalance in the Rabi frequency.  
The theory lines are obtained by leaving the Rabi frequency as a free parameter in one of the curves and scaling the other curves according to the MW power applied in the experiment.
The data demonstrates the compensation of the differential Zeeman shift around the field value of $B_\text{center} = 2.65$\,G.

\begin{figure}
\centerline{\includegraphics[width=.5\textwidth]{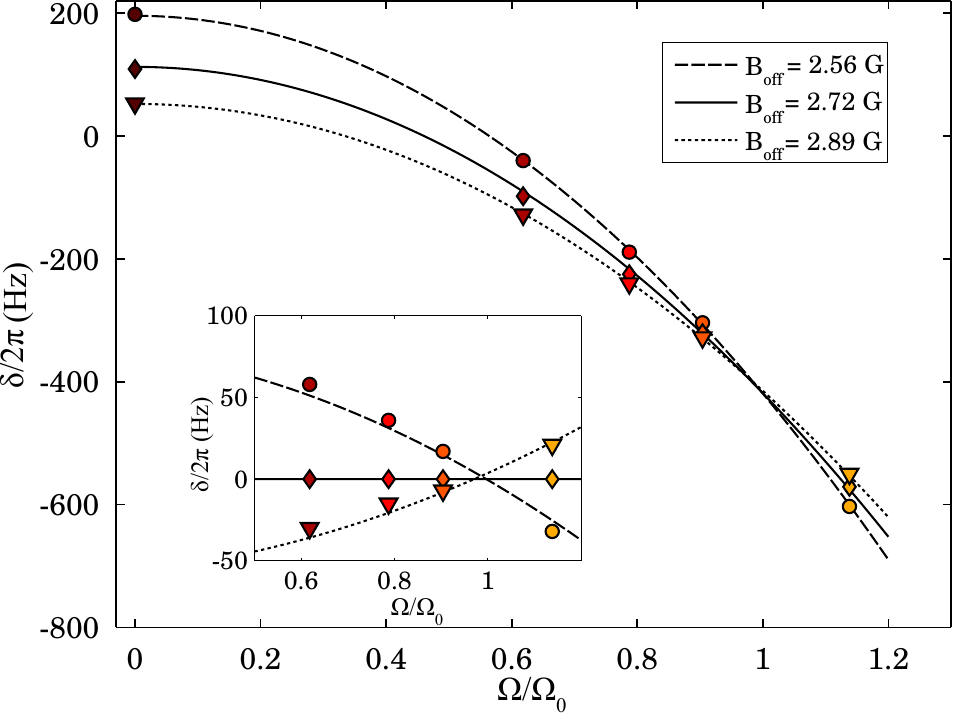}}
\caption{Frequency difference of the clock transition as a function of the Rabi frequency of the dressing for different magnetic fields. The data was extracted from the measurements in Fig.\ \ref{fig:results}. \textit{Inset}: Frequency difference for different offset fields with respect to the measurements at $B_\text{off} = 2.72$\,G as a function of the Rabi frequency. At the optimal Rabi frequency $\Omega_0$, the three curves show nearly identical frequencies, proving the cancellation of the differential Zeeman shift up to second order.}
\label{fig:freq_v_rabi}
\end{figure}

\begin{figure}
\centerline{\includegraphics[width=.5\textwidth]{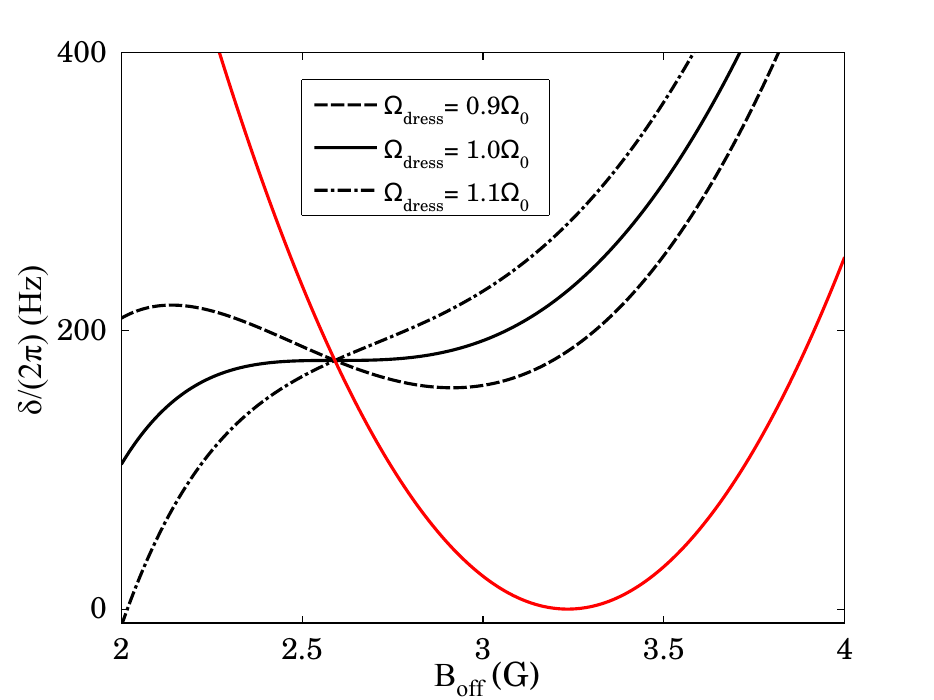}}
\caption{\textit{Double magic} dressing of the atomic clock transition, for which the dependence of the frequency both on the magnetic field and the Rabi frequency disappears around a field value of $B_\text{center}= 2.59$\,G. The calculation assumes a Rabi-frequency imbalance of $\sqrt{6}\Omega_{0,\sigma +}/\Omega_{0,\sigma -} = 1.25$, the obtained optimal parameters are $\Omega_0 = 2\pi\cdot 86.7$\,kHz, $\Delta_\text{dress}= -2\pi\cdot 309$\,kHz}
\label{fig:magic}
\end{figure}

The reduced sensitivity of the clock transition to magnetic field variations is shown in Fig.\ \ref{fig:freq_v_rabi}. 
Here we plot the measured frequencies and the theory curves for three different offset fields as a function of the Rabi frequency, as extracted from the values in Fig.\ \ref{fig:results}.
For the optimum Rabi frequency $\Omega_0$, all three curves show the same ac-Zeeman shift.
The inset in Fig.\ \ref{fig:freq_v_rabi} shows the frequency difference between the curves measured for the three offset fields with respect to the value $B_\text{off} = 2.72$\,G.
The three curves cross nearly at the same point, showing the strong suppression of the differential Zeeman shift over a field range of larger than 0.2\,G.
The analysis of the theory curves in Fig.\ \ref{fig:results} shows that it is possible to generate plateaus where the frequency differs by less than 0.1\,Hz over a magnetic field range of more than 100\,mG.
As visible in the inset of Fig. \ref{fig:results}, the measurement does not reach this accuracy. 
We estimate a frequency uncertainty of $\pm 5$\,Hz, based on the limited time between the Ramsey pulses and the uncertainty of the unstabilized microwave power.
The stability of the microwave Rabi frequency is expected to be the strongest limitation on the frequency stability. 
In order to reach the 0.1\,Hz range at the field point of 2.65\,G, a power stability on the order of $\Delta \Omega_\text{dress}/\Omega_\text{dress} \sim 10^{-4}$ would be required. 
For certain offset fields, however, it is possible to find solutions for Eq.\ \ref{eq:diff} where both the $B$-field dependency as well as the dependency on the the Rabi frequency $\Omega_\text{dress}$ disappear. 
An example for such a solution can be seen in Fig.\ \ref{fig:magic}: Here, we calculate that the transition frequency varies by less than $\pm 0.1$\,Hz over a range of 100\,mG around $B_\text{center} = 2.59$\,G.
At the center of the plateau, the frequency $\delta$ becomes independent of the Rabi frequency for a detuning of $\Delta_\text{dress} = -2\pi\cdot 309 $\,kHz.
In a range of $\pm 10$\,mG around $B_\text{center}$, a Rabi-frequency stabilization on the order of 1\% would be sufficient to reach a level of 0.1\,Hz stability. 
Such \textit{double magic} dressing enables the employment of this technique with on-chip microwave devices, where Rabi frequencies are inversely proportional to the distance to the chip.

Manipulation of the differential Zeeman shift can be used to decrease the frequency spread over the size of the cloud.
%
%For a cloud of $N =5\cdot10^4$ atoms at $T = 250\,$\,nK and $B_\text{off} = 2.65$\,G, the  frequency spread due to the inhomogeneity of the magnetic field without dressing is on the order of $\sigma_\mathrm{inh} \approx4$\,Hz, about an order of magnitude larger than the spread $\sigma_\mathrm{dens}$ caused by the inhomogeneous mean field interaction due to the density distribution in the trap \cite{Rosenbusch2009}.
%
For a cloud of $N =5\cdot10^4$ atoms at $T = 250\,$\,nK and $B_\text{off} = 2.65$\,G, the  standard deviation of the frequency distribution due to the inhomogeneity of the magnetic field without dressing is on the order of $\sigma_\mathrm{inh} \approx4$\,Hz, about an order of magnitude larger than the spread $\sigma_\mathrm{dens}$ caused by the inhomogeneous mean field interaction due to the density distribution in the trap \cite{Rosenbusch2009}.
%
%Dressing with a spatially homogeneous Rabi frequency can decrease $\sigma_\mathrm{inh}$ to a level smaller or equal to $\sigma_\mathrm{dens}$, leading to mutual compensation of the two effects.
%
Microwave dressing can be employed to decrease $\sigma_\mathrm{inh}$ to a level on the order of $\sigma_\mathrm{dens}$, thereby balancing the two effects and leading to a nearly homogeneous frequency over the size of the cloud.
For the parameters above and our trap, we calculate  the differential Zeeman shift can be engineered to cancel the collisional frequency shift down to a level of $\sigma \approx  2\pi\cdot 0.25$\,Hz.

In addition, our scheme can  be used to prepare states with  nearly arbitrary $\delta$ vs.\ $B_\text{off}$ curvatures around the desired field point, enabling one to suppress, enhance or spatially structure the differential Zeeman shifts.
Enhancing the $B$-field dependence could, for example, be used to counteract the strong mean field shifts in a Bose-Einstein condensate.
Further engineering of differential clock frequencies can be achieved by using multi-frequency microwave fields. 
This opens up new possibilities for microwave and radio-frequency dressing of atomic transitions, which has previously been used for trapping and manipulating of cold atoms \cite{Zobay2001,Colombe2004, Hofferberth2006, Lesanovsky2006, Heathcote2008} and the generation of state-dependent potentials \cite{Courteille2006, Boehi2009}.

\section{Conclusion}
In summary, we have shown both experimentally and theoretically that dressing of Zeeman sublevels in magnetically trapped atoms can render hyperfine transitions insensitive to magnetic field fluctuations around an arbitrary field value.
We have furthermore identified \textit{double magic} points, where the clock frequency becomes independent of the Rabi frequency.
Microwave dressing can be used to enhance the coherence time of quantum superposition states in arbitrary magnetic fields and for the creation of noise protected quantum memories.
The scheme is further applicable in atomic clock schemes in magnetically noisy environments or portable setups.

\section*{Acknowledgements}
The authors would like to thank Rainer Dumke and Thomas Judd for useful discussions.
This work was supported by the Deutsche Forschungsgemeinschaft (SFB TRR21) and the European Commission (FP7 STREP project ``HAIRS'').


%\bibliography{Literatur}

\end{document}